# Rendering graphene supports hydrophilic with non-covalent aromatic functionalization for transmission electron microscopy.

Radosav S. Pantelic [a,1], Wangyang Fu [b], Christian Schoenenberger [b], Henning Stahlberg [c]

(a) National Cancer Institute, 50 South Drive, Building 50, Room 4306, Bethesda, Maryland 20892
(b) Department Physik, University of Basel, Klingelbergstrasse 82, Basel CH-4056
(c) Center for Cellular Imaging and NanoAnalytics, University of Basel, Mattenstrasse 26, WRO-1058, Basel CH-4058

**Abstract**
Amorphous carbon films have been routinely used to enhance the preparation of frozen-hydrated transmission electron microscopy (TEM) samples, either in retaining protein concentration, providing mechanical stability or dissipating sample charge. However, strong background signal from the amorphous carbon support obstructs that of the sample, and the insulating properties of amorphous carbon films preclude any efficiency in dispersing charge. Graphene addresses the limitations of amorphous carbon. Graphene is a crystalline material with virtually no phase or amplitude contrast and unparalleled, high electrical carrier mobility. However, the hydrophobic properties of graphene have prevented its routine application in Cryo-TEM. This letter reports a method for rendering graphene TEM supports hydrophilic - a convenient approach maintaining graphene's structural and electrical properties based on non-covalent, aromatic functionalization.

Cryo-Transmission Electron Microscopy (Cryo-TEM) refers to the imaging of frozen-hydrated biological specimens by transmission electron microscopy (TEM), thereby preserving the high-order structure of protein complexes (and other biological samples) in a near-native state. Additional, thin (~2 nm) amorphous carbon films (spanning the otherwise freestanding vitreous ice) are routinely used to retain/promote mono-disperse sample concentrations [1]. Vitreous ice, which is between 2-100 times less rigid than amorphous carbon (according to measurements of bulk properties [2]) furthermore suffers from mechanical instability as beam induced radiolysis generates localised pockets of internalised pressure across the sample [3]. Hence, additional amorphous carbon films have also been used to improve mechanical stability (routinely with 2D protein crystals, for example) [4-7]. Inelastic interaction of the electron beam with the sample releases Auger and secondary electrons yielding net charge. Ions within the buffer that would otherwise transport/disperse charge are immobilised by vitrification, allowing the accumulation of localised charging. Immobile surface charges exert repulsive Coulomb forces, contributing to instability across the sample [2,8] and/or cause micro-lensing effects that induce beam hysteresis (further compromising imaging stability) [9-12]. Additional amorphous carbon films have been thought to discharge the otherwise insulated vitreous sample [2,4-6,8-11].

---

[1]To whom correspondence should be addressed: Email: pantelic@imbb.forth.gr

Contrary to the accepted rationale, the surface of thin, evaporated amorphous carbon films are electrically insulated, thus limiting any efficiency in dissipating sample charge [3,8]. Although electrical conductivity increases linearly with thickness in excess ~5.6 nm [8], increased bulk none the less fails to compensate for the semiconductor nature of the material at low temperatures (decreasing electrical conductivity) [5]. Furthermore, even at ~2 nm thickness amorphous carbon films introduce a strong background signal (both phase and amplitude contrast) that attenuates if not completely obstructs that of unstained molecules [13,14]. Graphene directly addresses the limitations of amorphous carbon films. The crystalline structure of pristine graphene demonstrates virtually no phase contrast down to 2.13Å [15]. At ~0.34 nm single-layer thickness [16], amplitude contrast (noise) from inelastic scattering within the support is also minimal [15]. Electrical conductivity is more than 6 orders of magnitude greater than that of amorphous carbon (converted to bulk units and assuming a thickness of 3.4 Å) [17-19], and remains unattenuated at low temperatures [20,21]. Hence, a growing interest in graphene TEM supports has emerged [6,12,14,15,22-31].

However, the widespread application of graphene in Cryo-TEM has been precluded by a lack of suitable (i.e. with minimal attenuation of crystalline structure, transparency and electrical mobility) and convenient (consistent & reproducible blotting, requiring no specialised equipment) methods for rendering graphene TEM supports hydrophilic. We report such a method, both convenient and maintaining crystalline structure, transparency and electrical properties (as demonstrated by Raman spectroscopy, Cryo-TEM and electrical measurements respectively).

Amorphous carbon supports are typically rendered hydrophilic by plasma etching and the subsequent implantation of –OH, -COOH and C=O groups (for example). However, carbon sputtering by incident ions destroys the crystalline graphene structure, degrading the material properties. Aromatic molecules such as pyrene have a strong affinity for graphene, stacking across the basal plane via stable π-π interactions [32] and include a variety of functionalities (e.g. -OH and -COOH) that have been used to create soluble dispersions of carbon nanotubes and graphite [33-38]. Here we demonstrate the preparation of frozen-hydrated biological samples across graphene TEM supports, pretreated by non-covalent (unlike covalent chemistries that disrupt underlying structure), aromatic functionalization with 1-PyreneCarboxylic acid (-COOH).

Graphene samples produced by chemical vapour deposition (CVD) were pre-etched and transferred to Si wafer (with 300 nm thermally grown $SiO_2$ layer)[39,40] for Raman and electrical measurements. After vacuum annealing at 300°C to remove all traces of Polymethyl methacrylate (PMMA), samples were left overnight (room temperature) in 1-PyreneCarboxylic acid (-COOH, Sigma-Aldrich art. # 391581) diluted to 100mM concentration in *N,N*-Dimethylformamide (DMF, Sigma art. 227056), then rinsed thoroughly with ethanol to prevent precipitate contamination upon drying.

Raman characterisation (WITec Alpha300R operated at room temperature with 532-nm laser excitation) before (black) and after (red) functionalization with 1-PyreneCarboxylic acid is shown in Fig. 1. Graphene structure is typically characterised by Raman peaks in the vicinity of ~1360 $cm^{-1}$, 1560 $cm^{-1}$ and 2700 $cm^{-1}$ ($I_D$, $I_G$ and $I_{2D}$ respectively, black curve). A distinctive up-shift of $I_G$ and $I_{2D}$ peaks by ~5.7 $cm^{-1}$ as well as increase of $I_G/I_{2D}$ ratio from 0.68 to 1.54, indicate p-type doping of the graphene by an electron-acceptor (such is 1-PyreneCarboxylic acid) after functionalization (red curve)[33]. Given the variance often encountered across individual graphene samples[41], datasets of 100 spectra were mapped across 10 x 10 $\mu m^2$ areas at defined positions before and after functionalization. On average, 100 spectra sampled across the same area demonstrated congruent $I_G$ and $I_{2D}$ shifts of ~3 $cm^{-1}$ and an increase in the average $I_G/I_{2D}$ ratio from 0.63 to 1.04. Pristine graphene may demonstrate minor structural disorder as indicated by weak $I_D$ amplitudes (as seen here even before any kind of functionalization, black curve), often in the vicinity of grain boundaries (for example). However, the minimal increase in $I_D/I_G$ ratio after functionalization from 0.09 to 0.17 (0.1 to 0.2 on average) further indicates a primarily non-covalent/non-destructive mode of functionalization (i.e. doping). What may be initially interpreted as a broadening of the D peak ($I_D$) or merging of $I_D$ and $I_G$ modes (red curve), after separate Raman experiments (data not shown) appears to be superimposed pyrene structure (grey arrows) as indicated by additional amplitude/peaks at 1234 $cm^{-1}$, 1387 $cm^{-1}$ and 1629 $cm^{-1}$. Doping with 1-PyreneCarboxylic acid maintains the crystalline structure from which graphene's unique properties are derived.

The electrical transfer characteristics of a graphene field-effect transistor (FET) before (black curve) and after (red curve) functionalization with 1-PyreneCarboxylic acid are shown in Fig. 2. Electron beam lithography was used to etch source/drain electrodes from Ti (5 nm) and Au (60 nm) bilayers. Electrical measurements were performed at a low source-drain voltage ($V_{sd}$) of 10-50 mV in the so-called "linear regime" using a Keithley 2600A source meter. The plot shows the electrical sheet conductance (G) measured as a function of the back gate voltage ($V_g$). A bipolar transistor characteristic is observed before functionalization (black curve), reflecting how the type of carriers in graphene can be continuously tuned from holes (p-type, descending left regime) to electrons (n-type, ascending right regime). The transition point (so-called charge neutral point, CNP) between the hole (p-type) and electron (n-type) regimes lies at the sheet conductance minimum, which in this case was at ~29 V ($V_g$, within expected ranges for monolayer graphene spanning an $SiO_2$ substrate[40]). The shape and gradient (descending) of the functionalized curve (red) with its expected CNP beyond 40 V ($V_g$) is indicative of a strong p-type doping, confirming such indications by Raman (Fig. 1). The high degree of p-type doping seen here increases the hole carrier concentration, elevating sheet conductance (G) and the effective dispersion of charge by the graphene. Similar would be expected in instances of strong n-type (electron-donor) doping also. Generally speaking, the conductance of graphene is expected to improve with elevated hole or electron concentration [42]. The conductance (central to the ability to discharge an otherwise electrically insulated vitreous sample) is proportional to the product of the graphene's carrier concentration and carrier mobility. The carrier mobility characterises how quickly an electron or hole carrier moves through the graphene. The calculated carrier mobility of the graphene FET before and after -COOH/Pyrene functionalization changes only sightly from 2800 cm$^2$/Vs to 2200 cm$^2$/Vs, many orders of magnitude greater than that reported for amorphous carbon ($< 10^{-3}$ cm$^2$/Vs) [43]. The results demonstrate that non-covalent functionalization with 1-PyreneCarboxylic acid, preserves the extraordinarily high carrier mobility and thus, excellent electrical properties of graphene, surpassing amorphous carbon by far.

In previous work we established the high transparency of graphene TEM supports compared to thin amorphous carbon films (~2 nm thickness) [13] and seek to qualify this optimized approach with comparable results. Graphene TEM supports were prepared as previously described [15]. We found it sufficient to incubate grids in 1-PyreneCarboxylic acid (Sigma-Aldrich art. # 391581) diluted to 100mM concentration in *N,N*-Dimethylformamide (DMF, Sigma art. 227056) for ~2 minutes. Before drying, grids were thoroughly rinsed in ethanol. Tobacco Mosaic Virus is a highly ordered helical virus, the periodic structure of which manifests 3rd and 6th order layer lines (apparent in the average power spectral density) at ~23 Å and ~11.5 Å respectively. The presence and signal to noise ratio's (SNR) of diffracted layer lines (including weaker orders) provides a relative indication of background transparency under realistic vitreous sample conditions. With no further preparation of the support after aromatic functionalization, 4 μl of Tobacco Mosaic Virus (TMV) diluted in buffer (50mM Tris-HC, 100mM MgCl$_2$, 1 mM EDTA, pH 7.7) to a final concentration of ~0.08 mg/ml was left to incubate on the grid for ~2 minutes before blotting (FEI Vitrobot IV, 1s blot, blot force -5, 20$^{\circ}$C, 100% humidity) and plunging into liquid ethane.

After functionalization with 1-PyreneCarboxylic acid, distinctive changes in the observable blotting characteristics (i.e. reduced surface tension) across the graphene support are apparent with thin aqueous films left spanning the grid (as observed by light microscopy). The presence of continuous blotting gradients trailing the peripheries of foil squares (Fig. 3(a) ) are as those found across conventionally treated amorphous carbon grids and indicate the minimal surface tensions requisite in blotting thin aqueous films for Cryo-TEM. Figure 3(b) shows vitrified TMV prepared across graphene, non-covalently functionalized with 1-PyreneCarboxylic acid. Imaging at -1.1 μm defocus, the theoretical zero-crossings of the CTF do not cancel the 3$^{rd}$ (~23 Å), 6$^{th}$ (~11.5 Å) and 9th (~7.7 Å) order layer lines. However, such close to focus imaging typically yields particularly low contrast images (especially so when imaging across even the thinnest amorphous carbon). Given the graphene's high transparency, individual TMV fibres are clearly discernible. Micrographs were normalized and average Fourier transforms were calculated from partially (25%) overlapping 512 x 512 pixel regions cropped along individual TMV fibres (a total of 313 micrographs were analysed, across different areas and grids). The power spectral density (PSD) of Fig. 3(b) (Fig. 3(c), calculated from 9 partially segments) demonstrates not only 3rd and 6th order layer lines with high SNR (3.75 and 2.4 respectively) but the emergence of weaker reflections (1st, 2nd, 4th & 5th order). On average SNR's of 3.1 and 1.19 for 3rd and 6th order layer lines (respectively, calculated from 5 overlapping segments) are comparable to previous results [13]. The strong background signal (both phase and amplitude) contributed by amorphous carbon films, regardless of imaging parameters, obstructs that of low contrast vitrified biological molecules. Our results indicate no apparent introduction of background signal following functionalization with 1-PyreneCarboxylic acid.

Despite Cryo-TEM's growing interest in graphene supports, the fundamental issue of hydrophobicity has been largely if not completely neglected. The inert and hydrophobic nature of graphene prevents the blotting of thin aqueous films amenable to vitrification. Conventional plasma treatments render amorphous carbon supports hydrophilic through disruption of underlying structure and implantation of oxygen and hydrogen (for example) ions (covalent functionalization). The unique imaging properties of graphene are intrinsic to its highly ordered crystalline structure and are severely attenuated/degraded by basal plane defects such as those introduced through covalent functionalization. Hence, plasma treatments are by definition not suited to the preparation of graphene supports. Although the supporting amorphous Quantifoil may somewhat aid in retaining small amounts of buffer (i.e. applying sample to the back of the grid), continuous aqueous films prove often elusive as closer examination of the graphene often reveals only partially vitreous areas where high surface tension across the graphene has constrained buffer to small droplets amidst the otherwise completely evaporated support. Deposition of an additional amorphous carbon layer across the graphene may be amenable to plasma treatments but contributes significant background noise. Any type of amphiphilic monolayer (i.e. detergents, PEG and other polymers) assembled across the graphene would yield similar degradation of signal. In continuation from previous work we report a simple and convenient method based on non-covalent aromatic functionalization with 1-PyreneCarboxylic acid, by which graphene TEM supports may be rendered hydrophilic whilst maintaing the key, aforementioned properties of graphene. The pyrene based molecule is preferable to others with similar chemistries such as phenol and phloroglucinol (-OH) as the larger aromatic molecule attaches with greater stability (improving consistency/reliability).

We have noted that the efficacy of this method is dependent upon the pristine state of the graphene support, since significant contamination (either from etching or transfer residues) interferes with π-π electron interactions (π-stacking) [44]. Exhaustive etching and cleaning during transfer of the graphene TEM supports appears to adequately address this issue [15]. Studies by single particle analysis rely upon the random orientation of complexes in order to sufficiently sample all angular projections. Hence, explicit protein functionalization has not been considered since interaction with protein amine groups [45] (for example) may introduce preferential orientation of complexes. However, various pyrene chemistries (i.e. aminopyrene) provide a pathway to specific surface affinities that may be of further interest in future applications (i.e. in-situ affinity purification of EM samples).

With virtually no attenuation of the graphene's transparency or electrical mobility, we anticipate the sheer simplicity of this approach will facilitate the wider use of graphene in conventionally equipped TEM labs for the optimisation of Cryo-TEM samples.

**Acknowledgements**
This work was completed in equal contribution by Dr Radosav Pantelic and Dr Wangyang Fu. Dr Radosav Pantelic wishes to thank Dr Sriram Subramaniam (National Cancer Insitute/National Institutes for Health) as sponsor of his fellowship at the National Institutes for Health (NIH) and access to TEM resources. Prof. Henning Stahlberg appreciates the support of the Swiss National Science Foundation (NCCR Nano and Structural Biology) and Swiss Initiative for Systems Biology (SystemsX.ch, RTD CINA). Prof. Christian Schoenenberger and Dr Wangyang Fu gratefully acknowledge Funding from the Swiss Nationalen Forschungsschwerpunkte (NFS), Nano-Tera.ch, Swiss Nanoscience Institute and ESF-Eurographene.

**Figure 1. Raman spectroscopy before/after functionalization with 1-PyreneCarboxylic acid.** Comparison of Raman spectra before (black) and after (red) functionalization indicate maintenance of characteristic graphene structure with distinctive up-shift of $I_G$ and $I_{2D}$ modes (~5.7 cm$^{-1}$) and increase of $I_G/I_{2D}$ ratio (from 0.68 to 1.54) further indicating p-type doping. The non-covalent/non-destructive nature of the functionalization is further reflected by the only slight increase in $I_D/I_G$ ratio before (0.09) and after (0.17) treatment. We also see the superimposition of pyrene structure as additional peaks (grey arrows) at 1234 cm$^{-1}$, 1387 cm$^{-1}$ and 1629 cm$^{-1}$.

**Figure 2. Electrical transfer curves before/after 1-PyreneCarboxylic acid functionalization.** The electrical transfer curves (conductance (G) as a function of back gate voltage ($V_G$)) of a graphene FET before (black) and after (red) functionalization with 1-PyreneCarboxylic acid. Before treatment, a bipolar transistor characteristic is observed (arrows indicate the relationship between doping and gradient, the position of the CNP). However, after functionalization the shape of the transfer curve changes to that of a p-type doping (note gradient and shape of red curve, in correspondence with the shifting of Raman modes in Fig. 1).

**Figure 3. Electron microscopy of frozen-hydrated TMV across graphene.** Grid squares demonstrate continuous blotting gradients at their peripheries (a) indicating the minimal surface tensions (hydrophilic surface) requisite in blotting thin aqueous films. TMV is clearly discernible (b) despite the additional support and relatively close to focus (-1.1 µm) imaging. The calculated average power spectral density in panel (c) shows not only the 3rd and 6th-order layer lines in striking contrast (3.75 and 2.4 respectively) but also the emergence of 1st, 2nd, 4th & 5th order reflections. TMV was imaged at 300 keV low dose (~25 e/Å$^2$), recorded at 47,000x magnification (1.4 Å pixel size) and 1.1s exposure time.

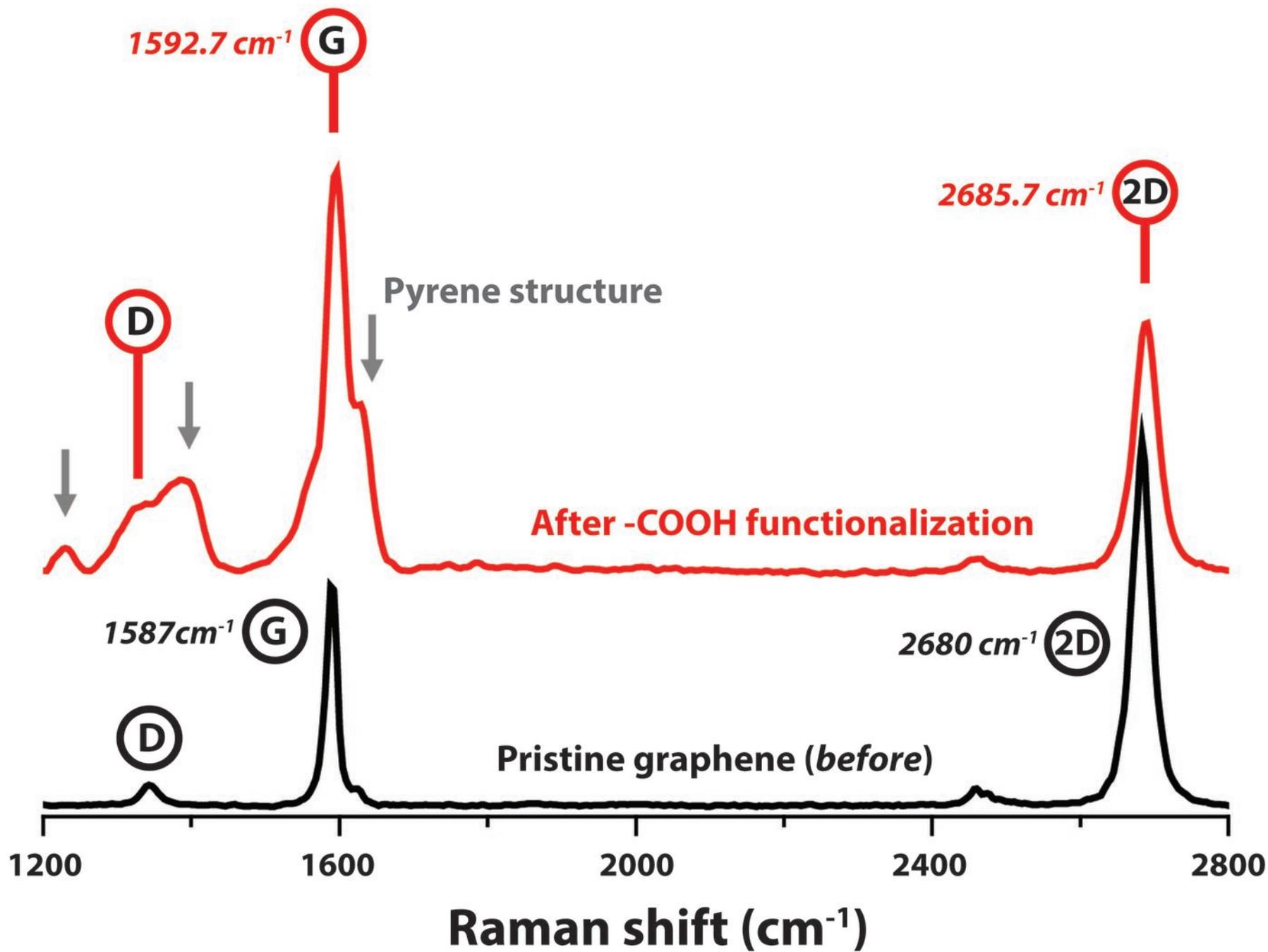

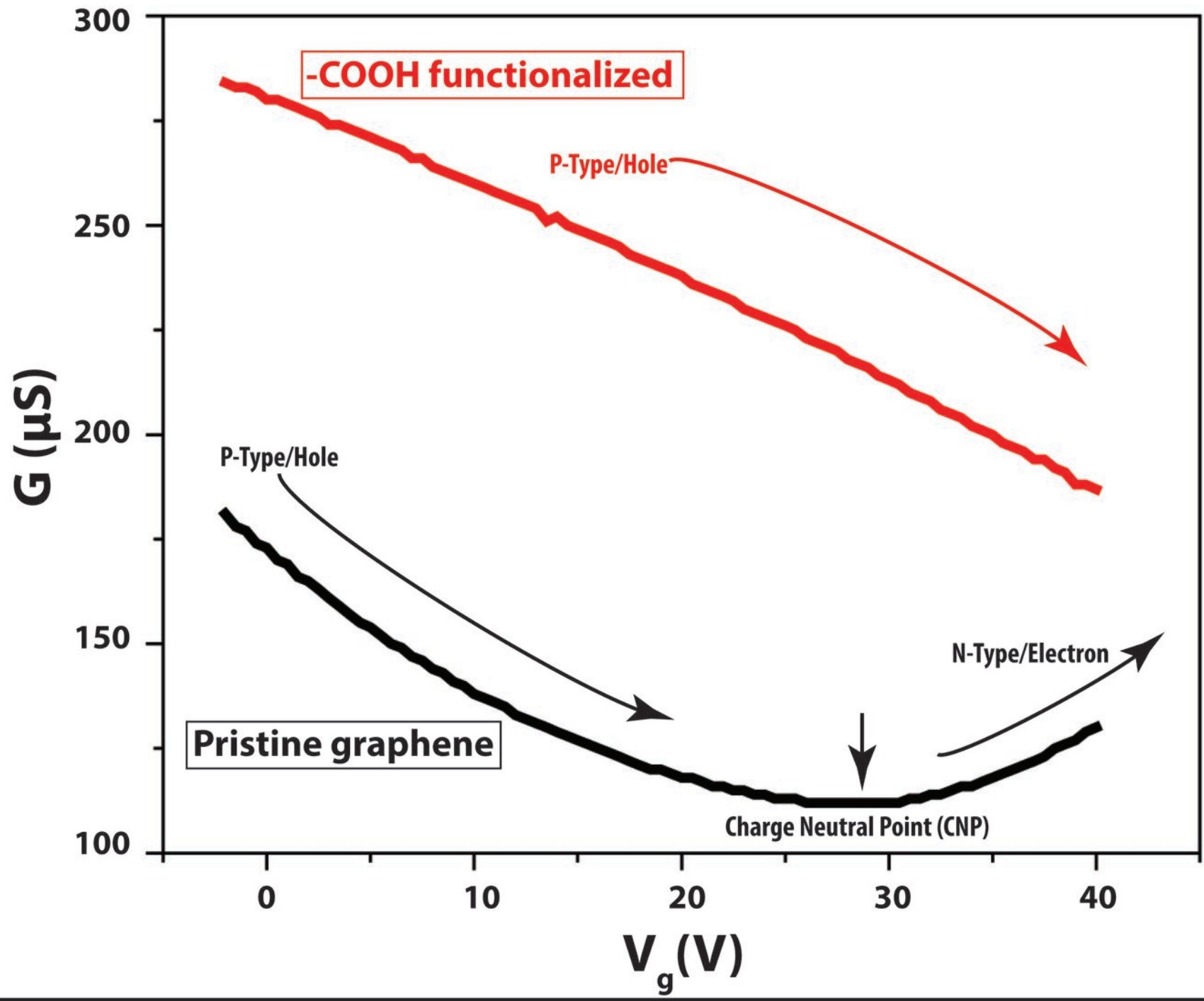

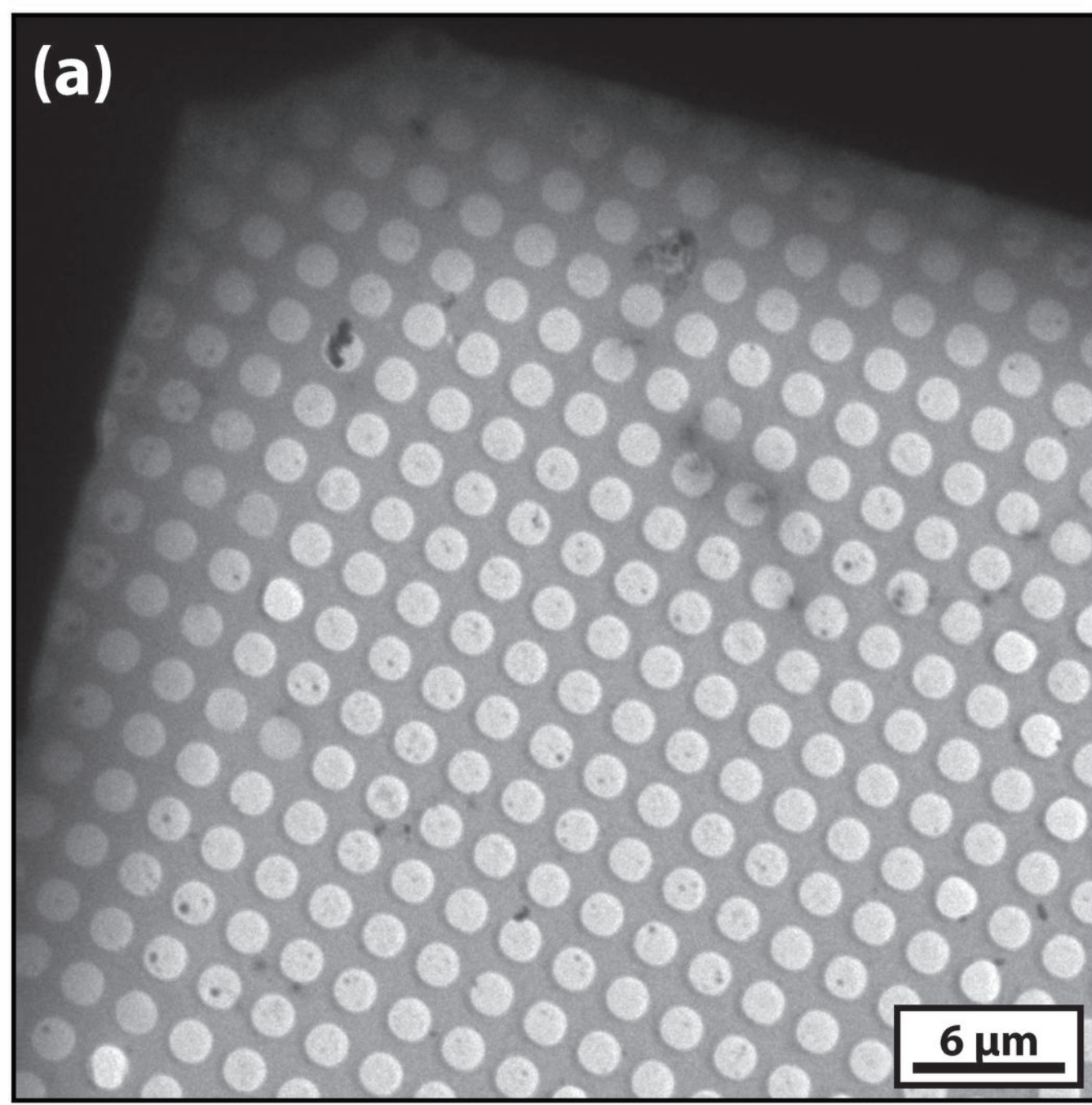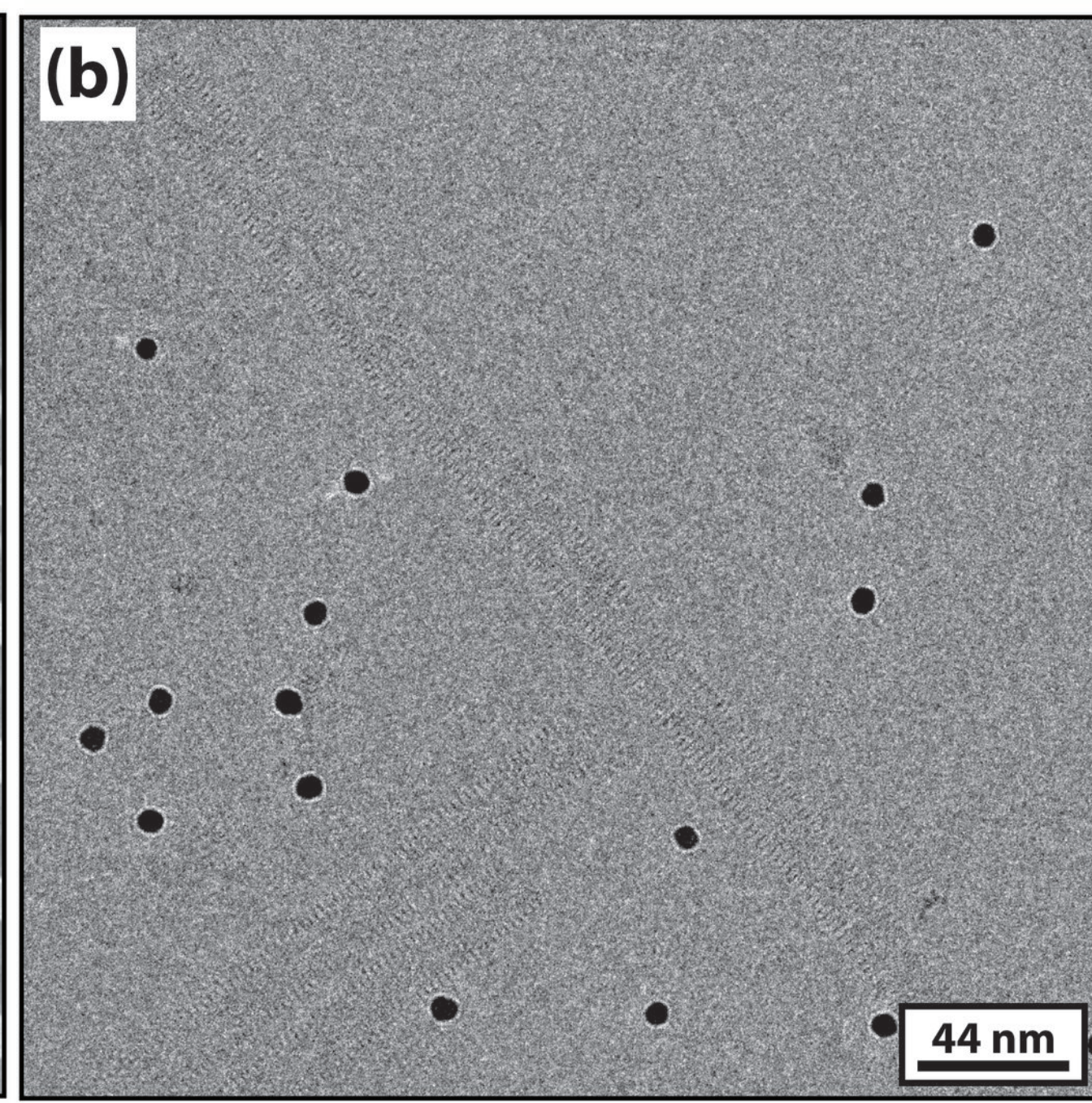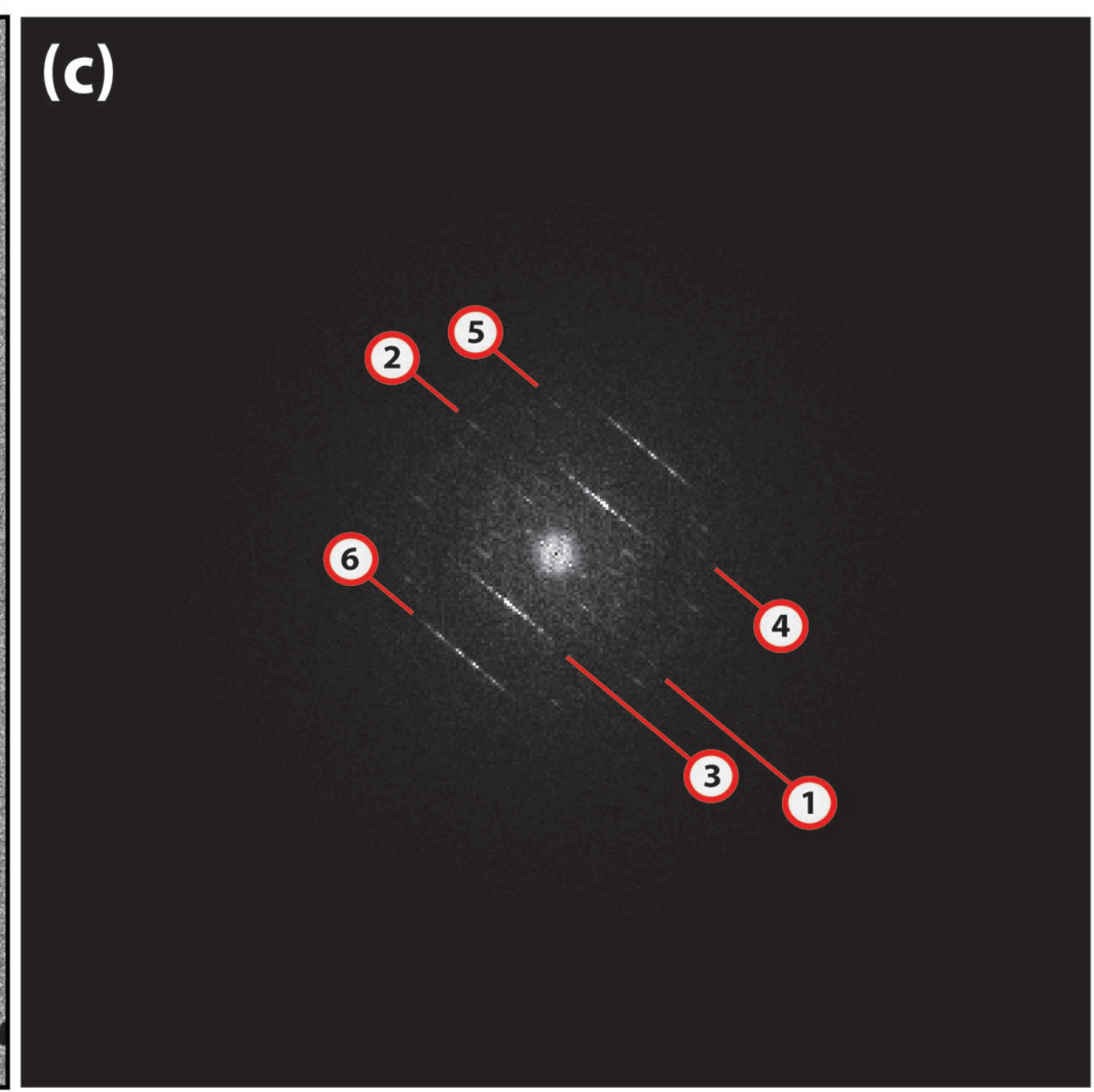